\documentclass[12pt,preprint]{aastex}

\usepackage{lscape}

\begin{document}

\title{Tagging the chemical evolution history of the Large Magellanic 
Cloud disk
\footnote{Based on observations collected at the ESO-VLT at Cerro Paranal (Chile) under the program 080.D-0368(A).}}

\author{Emilio Lapenna}
\affil{Dipartimento di Astronomia, Universit\`a degli Studi di Bologna, Via Ranzani, 1 - 40127 Bologna, ITALY}
\email{emilio.lapenna2@unibo.it}

\author{Alessio Mucciarelli}
\affil{Dipartimento di Astronomia, Universit\`a degli Studi di Bologna, Via Ranzani, 1 - 40127 Bologna, ITALY}
\email{alessio.mucciarelli2@unibo.it}

\author{Livia Origlia}
\affil{INAF - Osservatorio Astronomico di Bologna, Via Ranzani, 1 - 40127 Bologna, ITALY}
\email{livia.origlia@oabo.inaf.it}

\and

\author{Francesco R. Ferraro}
\affil{Dipartimento di Astronomia, Universit\`a degli Studi di Bologna, Via Ranzani, 1 - 40127 Bologna, ITALY}
\email{francesco.ferraro3@unibo.it}

\begin{abstract} 
We have used high-resolution spectra obtained with the multifiber facility FLAMES at the Very Large Telescope of the 
European Southern Observatory to derive kinematic properties and chemical abundances of Fe, O, Mg and Si for 89 
stars in the disk of the Large Magellanic Cloud (LMC). 
The derived metallicity and [$\alpha$/Fe], obtained as the average of O, Mg and Si abundances, allow us to 
draw a preliminary scheme of the star formation history occurred in this region of the LMC. The derived metallicity distribution 
shows two main components: one component (comprising $\sim$ 84$\%$ of the sample) is peaked at [Fe/H] = --0.48 dex and it shows 
an [$\alpha$/Fe] ratio slightly under solar ([$\alpha$/Fe] $\sim$ --0.1 dex). 
This population was probably originated by the main star formation event occurred 3--4 Gyr ago (possibly triggered by tidal capture of 
the Small Magellanic Cloud). The other component (comprising $\sim$ 16$\%$ of the sample) is peaked at [Fe/H] $\sim$ --1 dex and it 
shows an [$\alpha$/Fe] $\sim$ 0.2 dex. This population  was probably generated during the long quiescent epoch of star formation in 
between the first episode and the most recent bursts.
Indeed, in our sample we do not find stars with chemical properties similar to the old LMC globular clusters nor to the iron-rich 
and $\alpha$-poor stars recently found in the LMC globular cluster NGC~1718 and predicted to be also in the LMC field, thus suggesting 
that both these components are small ($<$ 1$\%$) in the LMC disk population.
\end{abstract}


\section{Introduction}

Despite several years of studies, the chemical evolution history of the Large Magellanic Cloud (LMC) as well its star 
formation history (SFH) is still poorly understood. The LMC has experienced a complex SFH due to the 
interaction occurred both with the Small Magellanic Cloud (SMC) and the Galaxy \citep{bekki05}. A clear evidence 
of this quite complex evolution can be recognized in the LMC star cluster system, 
characterized by a wide range of age and metallicities. 
The study of these cluster stellar populations reveals the existence of at least three 
components: the old ($\sim$ 13 Gyr, \citealt{brocato96}, \citealt{olsen98}) metal-poor component ([Fe/H] $<$ --1.5 dex, 
\citealt{olszewski91}, \citealt{grocholski06}, \citealt{m10}) probably formed during the first episode of star formation 
(SF), the intermediate age component ($\sim$ 1--3 Gyr, \citealt{gallart03}, \citealt{ferraro04}) which is the dominant one, 
and a young component ($\lesssim$ 1 Gyr, \citealt{brocato03}, \citealt{m11}) that include the last formed clusters. 

The SFH of field stellar populations is less known. \citet{smecker02} found that the dominant stellar population in the central 
bar was formed between  4-6 Gyr and 1-2 Gyr ago. This result was substantially confirmed by the simulations of \citet{bekki05} who 
found that the formation of young stellar populations in the LMC bar is associated with efficient SF in the last few Gyr (in particular, 
$\sim$ 2 Gyr ago). 
A comprehensive spectroscopic study of field populations was performed by \citet{cole05} measuring the infrared CaII triplet of 
373 giant stars located around the center of the bar. They found a distribution peaked at [Fe/H] $\sim$ --0.4 dex with a low metallicity 
tail reaching [Fe/H] $\lesssim$ --2.0 dex. \citet{carrera08} have measured the infrared CaII triplet in four fields at different 
radial distances ($\sim$ 3$^{\circ}$, 5$^{\circ}$, 6$^{\circ}$ and 8$^{\circ}$, that is between 2.6 and 7 kpc) from the center of LMC. 
They found an average [Fe/H] $\sim$ --0.5 dex for the closest fields, a lower [Fe/H] $\sim$ --0.8 dex was found only in the more 
distant one. A detailed spectroscopic analysis of 59 giant stars located in the inner disk at $\sim$ 1.2 kpc from the center was 
performed by \citet{pompeia08}, who found a distribution peaked at [Fe/H] $\sim$ --0.75 dex.

This paper is part of a project devoted to investigate the kinematic and chemical properties of the stellar populations 
in the LMC through the use of high-resolution spectra, able to provide accurate information about the kinematics, the metallicity 
and the chemical abundance of individual elements of these stars. Previous papers of the project have discussed the properties 
of old \citep{m09,m10}, intermediate-age 
\citep{ferraro06,m08} and young \citep{m11,m12} globular clusters (GC) in the LMC. This paper is the first of the project dedicated to 
the kinematic and chemical characterization of the LMC field stellar populations: we present chemical patterns for a sample of 89 
Red Giant Branch (RGB) stars members of the LMC and located in the field around the old metal-poor GC NGC~1786.
This represents to date the largest sample of field giants in the LMC for which high-resolution spectra have been obtained.


\section{Observations}

We have observed 91 stars located in the region surrounding the old GC NGC~1786 at $\sim$ 2$^{\circ}$ NW from 
the center of the LMC \citep{kim98}. The spectra have been acquired with the multi-object spectrograph FLAMES \citep{pasquini02} 
at the Kueyen ESO-VLT telescope.

The spectroscopic targets have been selected by using the near-infrared (J, H and K bands) photometric catalog, obtained 
by combining the SOFI catalog for the inner $\sim$ 2.5 arcmin, corresponding to the area covered by the GC NGC~1786 
(see \citealt{m10} for details) and the Two Micro All Sky Survey (2MASS) catalog for the outermost region, in order to sample 
the surrouding field population. The final catalog was placed onto the 2MASS photometric and absolute astrometric system by following the standard procedure 
used in \citet{ferraro04}. Photometric uncertainties are of about 0.01-0.02 and 0.03-0.05 for SOFI and 2MASS target, respectively.
The targets have been selected along the RGB and in the magnitude range $K_{0}$ $\sim$ 12.3--14.0, in order to reach a 
sufficient SNR ($>$ 30-40) and excluding stars brighter than the magnitude level \citep[$K_{0}\sim$12.3, see][]{cioni06} of the 
RGB Tip of the intermediate-age LMC population.

Observations consists of a series of 45-minute-long exposures obtained with HR11 (5600--5840 $\rm\mathring{A}$ and R = 24200) 
and HR13 (6120--6400 $\rm\mathring{A}$ and R = 22500) gratings by using the GIRAFFE/MEDUSA configurations. The data reduction 
was performed with the GIRAFFE ESO pipeline that includes bias subtraction, flat fielding, wavelength calibration and spectrum extraction. 
Individual stellar spectra have been cleaned from the sky contribution, by subtracting the corresponding median sky spectrum. Finally,
multiple spectra of each target have been coadded, reaching a SNR per pixel of $\sim$ 40 in the faintest stars and up to $\sim$ 100 
in the brightest ones.
Identification number, right ascension and declination of each target are listed in Table \ref{taba}.


\section{Kinematic and chemical analysis}

Radial velocities ($v_{r}$) have been measured by means of the DAOSPEC code \citep{stetson08}. For each star, the spectra from 
the two different gratings have been analysed independently and the derived $v_{r}$ averaged together by using the individual 
uncertainty as weight. Typical internal errors (computed as $\sigma$/$\sqrt{N_{lines}}$) are $\sim$ 0.15--0.20 km s$^{-1}$. 
Finally, we applied the heliocentric correction to each $v_{r}$.

The chemical analysis has been performed by using the suite of codes developed by R.L.Kurucz (see \citealt{sbordone04}) aimed 
to computing abundances from the observed equivalent widths (EWs) and synthetic spectra. The model atmospheres were computed 
with the ATLAS9 code, assuming plane-parallel geometry, local thermodynamical equilibrium for all species and without the inclusion of the 
approximate overshooting in the computation of the convective flux. The ATLAS9 model atmospheres were calculated with the new set of 
Opacity Distribution Functions by \citet{castelli04}.

The line list has been selected starting from the Kurucz/Castelli data\footnote{http://wwwuser.oat.ts.astro.it/castelli/linelists.html} 
and updated with recent laboratory data from VALD and NIST databases. We included only transitions with theoretical/laboratory 
atomic data and checked against spectral blendings through the inspection of suitable synthetic spectra convolved at the GIRAFFE resolution.
In particular, we adopted for the Fe lines the atomic data from the critical compilations by \citet{fmw} and \citet{fw06}, that represent 
the most updated dataset of theoretical/laboratory log~gf for the iron lines.

For some unblended transitions for which theoretical/laboratory oscillator strengths are not available (or for those lines not well reproduced in the solar 
spectrum) we derived astrophysical oscillator strengths (labelled as SUN in Table \ref{tabb}) by using the solar flux spectra of \citet{neckel} 
and the model atmosphere for the Sun computed by F. Castelli\footnote{http://wwwuser.oat.ts.astro.it/castelli/sun/ap00t5777g44377k1odfnew.dat} 
adopting the solar abundances of \citet{grevesse98}. We estimated for these oscillator strengths an accuracy $<$ 15\%, according to the uncertainties 
in the line profile fitting and in the continuum placement.
We decide to include in our final linelist some Fe~I transitions listed by \citet{fmw} and \citet{fw06} with high quoted uncertainties 
after the verification that these lines are well reproduced in the solar spectrum of \citet{neckel} and providing an iron abundance 
within $\pm$ 0.1 dex from the value of \citet{grevesse98}.

Concerning the van der Waals damping constants, we adopted, whenever possible, the damping values by \citet{barklem00}, while for the 
other transitions the van der Waals constants were calculated according to \citet{castelli05}.
The complete line list used is available in Table \ref{tabb}, including wavelength, element code, oscillator strength (and 
corresponding accuracy), lower excitational potential and reference source.

The atmospheric parameters have been derived as follows:

{\sl (1)}~A preliminary estimate of $T_{\rm eff}$ for each star was obtained from photometric $(J-H)_{0}$ and $(J-K)_{0}$ colors 
adopting an average color excess of $E(B-V)$ = 0.12 \citep{persson83}, the extinction law by \citet{rieke85} and the color-temperature 
calibrations by \citet{alonso99} based on the Infra-Red Flux Method. 
The 2MASS magnitudes of our targets have been converted in the TCS photometric system (where the \citet{alonso99} relations are defined) 
by using the transformations provided by \citet{carpenter}. 
We used the photometric $T_{\rm eff}$ as {\sl first-guess values}  
and then we refined them by imposing the lack of any trend between Fe I abundances and the excitation potential $\chi$. 
Basically, we find a good agreement between the photometric and spectroscopic $T_{\rm eff}$, with an average 
difference of $T_{\rm eff}^{spec}$-$T_{\rm eff}^{phot}$=~139 K ($\sigma$ = 173 K).
Finally, the spectroscopic $T_{\rm eff}$ have been adopted for the following analysis. Typical uncertainties in the spectroscopic 
temperatures (calculated according to the slope in the plane A(Fe~I) vs $\chi$) ranging from $\sim$80 up to $\sim$130 K.
 
{\sl (2)}~Microturbulent velocities have been derived by requiring the lack of any trend between Fe I abundances and 
the reduced EW (defined as $\lg(EW/\lambda$). Typical uncertainty in the $v_t$ determination is 0.1-0.2 km/s.

{\sl (3)}~Surface gravities have been estimated from the photometry, because the classical method to infer log~g
from the comparison between Fe~I and Fe~II lines cannot be applied, due to the small number of available Fe~II lines in our spectra.
Surface gravities have been derived from the Stefan-Boltzmann relation, using for each target the corresponding spectroscopic $T_{\rm eff}$.
We adopted the distance modulus $(m-M)_{0}$ = 18.50 \citep{alves04} and a typical evolutionary mass (1.5 $M_{\odot}$)
for all the targets derived from a BaSTI \citep{pietrinferni04} isochrone with age of 2 Gyr, Z=~0.008 
and solar-scaled chemical mixture, according to the values 
estimated for the bulk of the LMC disk population by \citet{bekki05} and \citet{cole05}.
The luminosities are derived from the $M_{K}$-band magnitude using the bolometric corrections of \citet{buzzoni10}. 
Internal errors in the derived log~g (calculated by taking into account the uncertainties in the spectroscopic $T_{\rm eff}$, mass and luminosity) are of about 0.1-0.15 dex.
The photometric estimates of temperatures and gravities together with those obtained from spectroscopy are listed in Table \ref{tab1}.

EWs were measured with DAOSPEC. The abundances of Fe and Si have been obtained from the EWs.
O abundances have been derived from the forbidden line at 6300.31 $\rm\mathring{A}$ by employing spectral synthesis technique.
The abundances have been determined from $\chi^2$-minimization between the observed spectrum and a grid of 
synthetic spectra computed with the corresponding atmospheric parameters and convolved at the GIRAFFE resolution. 
Mg abundances have been derived with the same technique
by using the strong line 5711.09 $\rm\mathring{A}$; we decide to exclude the 
Mg triplet at $\sim$ 6318 $\rm\mathring{A}$ because these weak lines are located on the red wing of a close auto-ionization Ca line. This produce a decrease of the continuum level which was not easily measurable with DAOSPEC.
The final abundances are expressed in the usual spectroscopic formalism (logarithmic scale in unit of solar abundances) 
and adopting the solar values from \citet{grevesse98} except for oxygen, for which we used the value by \citet{caffau10}.
All the abundance ratios of our sample are listed in Table \ref{tab2}.


\section{Abundance uncertainties}

For each abundance ratio we take into account two source of uncertainties, the internal error relative 
to the EW measurements and that arising from the uncertainties in the atmospheric parameters.

Uncertainties for each abundance ratio due to the uncertainties in the EW measurements
have been estimated (for each star) by dividing the line-to-line scatter by the square root of the number of used lines. 
DAOSPEC provides also for each measured spectral line an error computed from the standard deviation 
of the local flux residuals (representing a 68\% confidence interval on the derived value of the EW). 
These uncertainties have been used only in the computation of the slopes in the A(Fe~I) vs $\chi$ and 
A(Fe~I) vs EWR planes (see Section 3).

Recently, \citet{venn12} pointed out that the internal EW errors provided by DAOSPEC are under-estimated of a 
factor of 2 with respect to those derived with the classical formula by \citet{cayrel88}, probably due 
to the pixel correlation during the wavelength calibration. We checked this effect in our data, assuming 
the FWHM estimated by DAOSPEC and the proper SNR for each spectrum. On average, we found a general 
agreement between the two sets of uncertainties (within $\pm$ 10-20\%) but not a systematic effect.
However, the DAOSPEC errors provide an internal ranking among the measured lines in a given star and 
possible systematic under/over-estimates do not afflict our results.

For O and Mg, for which we derived the abundances from spectral synthesis of only one line each, 
the uncertainty in the measurement has been estimated by resorting to Monte Carlo simulations. 
We study some stars as representative of the different SNR of our targets. We injected Poisson noise into the best-fit 
synthetic spectrum according to the SNR of the observed spectrum and we repeated the fit procedure; 
for each line we performed a total of 500 Monte Carlo events, assuming as uncertainty in the 
fitting procedure 1$\sigma$ of the abundances distribution. 
For the O line (falling in the HR13 grating), the derived uncertainties are of 0.04, 0.03 and 0.02 dex at SNR=~30, 50 
and 80 respectively, while 
for the Mg line (falling in the HR11 grating) we obtained uncertainties of 0.13, 0.09 and 0.04 dex for 
SNR=~30, 40 and 70.

We estimated also the uncertainty due to the continuum location, by using the root mean square of the 
residual provided by DAOSPEC, after removing all the fitted lines from the observed spectrum. For our stars 
the typical dispersion of the residual spectra is of $\pm$ 3-4\%, corresponding to an error in the abundances 
less than 0.1 dex. Keeping in mind that DAOSPEC estimates a global continuum along the entire spectrum 
(and not a local continuum for each individual transition), this uncertainty 
can be considered as a systematic errorbar for all the lines in a given spectrum.

The uncertainty arising from the atmospheric parameters has been computed following the approach 
described by \citet{cayrel04}. 
The usual method to derive the errors due to the stellar parameters is to vary one only parameter each time, keeping 
the other ones fixed and finally adding in quadrature the derived abundance variations. Obviously, this method 
does not take into account the correlations among the atmospheric parameters, providing only a conservative 
value for the uncertainty. In our case, 
$T_{\rm eff}$ and $v_t$ have been optimized spectroscopically (and log~g has been derived 
according to the spectroscopic temperature), thus all the parameters are not independent each other. 
For each star, the temperature has been varied by $\pm$ 1$\sigma_{T_{\rm eff}}$, because the uncertainty in $T_{\rm eff}$ 
dominates the derived abundances as pointed out by \citet{cayrel04}. 
Then, we repeated the optimization procedure described above by keeping the temperature fixed and deriving 
new values for log~g and $v_t$. The advantage of this method is to naturally take into account the covariance 
terms among the parameters \citep[see also][]{shetrone03}.
The procedure has been repeated independently for each star by considering the corresponding $T_{eff}$ uncertainty 
computed from the error of the slope in the A(Fe)---$\chi$ plane. Table \ref{tab3} lists the results of this procedure 
for a representative star of our sample: the second column shows the final uncertainty for each abundance ratio, 
while the other columns are the results obtained with the usual approach to vary independently each parameter. Last column 
of Table \ref{tab3} is the sum in quadrature of the uncertainties due to one only parameter: these values are 
larger than those obtained with the \citet{cayrel04} approach because the covariances among the parameters is neglected.
Typical abundance uncertainties due to the atmospheric parameters are 
of the order of [Fe/H] = $\pm$ 0.03 dex, [O/Fe] = $\pm$ 0.04 dex, [Mg/Fe] = 
$\pm$ 0.06 dex and [Si/Fe] = $\pm$ 0.09 dex.

Finally, the total internal error for each abundance ratio was obtained by adding in quadrature the error 
associated to EW measurements and atmospheric parameters. For [Fe/H] it turns out to be of the order of $\sim$ 
0.04-0.05 dex because of the large number of measured lines.


\section{Results}

The heliocentric radial velocity distribution of the stars in our sample is shown in the left panel of Figure~\ref{histogram}. 
Stars with radial velocity in the range 170 $\leq$ $v_{r}$ $\leq$ 380 km s$^{-1}$ are considered LMC members, according to 
\citet{zhao03}: only 2 stars with $v_{r}\sim$100 km s$^{-1}$ have been excluded (likely belonging to the Galaxy). The mean velocity 
of the sample is $v_{r}$=~259.3 km s$^{-1}$ ($\sigma_{v}$ = 33.9 km s$^{-1}$), in good agreement with previous measurements in other 
samples of LMC disk by \citet{cole05}, \citet{carrera08} and \citet{pompeia08}.

The metallicity distribution of our sample is shown in the right panel of Figure~\ref{histogram}. The entire sample has an average 
[Fe/H] = --0.58 dex ($\sigma_{\rm[Fe/H]}$ = 0.25 dex). Also, two main components can be distinguished:

\begin{enumerate}
\item A principal component (hereafter LMC-R) comprising $\sim$ 84$\%$ of the observed sample with [Fe/H] $>$ --0.7 dex, peaked at 
[Fe/H]= --0.48 dex and with a quite small dispersion ($\sigma_{\rm[Fe/H]}$ = 0.13 dex).
\item A secondary component (hereafter LMC-P), comprising $\sim$ 16$\%$ of the sample, peaked at [Fe/H] = --1.06 dex 
($\sigma_{\rm[Fe/H]}$ = 0.18 dex) with an extended tail up to [Fe/H] $\sim$ --1.5 dex.
\end{enumerate}

The peak of the metallicity distribution of LMC-R (see Fig.~\ref{mult_hist}, a) is in good agreement with those obtained by 
\citet{cole05} ([Fe/H] = --0.45 dex, $\sigma_{\rm[Fe/H]}$ = 0.31 dex) (see Fig.~\ref{mult_hist}, b) from the central bar and 
\citet{carrera08} ([Fe/H] = --0.47 dex ($\sigma_{\rm[Fe/H]}$ = 0.30 dex), [Fe/H] = --0.50 dex ($\sigma_{\rm[Fe/H]}$ = 0.44 dex) 
and [Fe/H] = --0.45 dex ($\sigma_{\rm[Fe/H]}$ = 0.31 dex) for the fields located at $\sim$ 3$^{\circ}$, 5$^{\circ}$ and 6$^{\circ}$ 
at north of LMC center, respectively). A distribution peaked at a slightly lower value ([Fe/H] = --0.75 dex, $\sigma_{\rm[Fe/H]}$ = 0.23 dex) 
was found by \citet{pompeia08} (see Fig.~\ref{mult_hist}).

Figure~\ref{omgsi} shows the behaviour of O, Mg and Si, as a function of [Fe/H]. We plotted for comparison the abundance 
ratios measured in other environments: the Galaxy \citep[including data for thin disk, thick disk and halo, by ][]{venn04, reddy06},
the LMC disk \citep{pompeia08}, the nearby dwarf spheroidals \citep{shetrone01, shetrone03,letarte,lemasle,venn12}, 
the Sagittarius dwarf galaxy \citep{sbordone07} and the old and intermediate-age LMC GCs \citep{johnson06, m08, m09, m10}\footnote{We excluded from this comparison the LMC GCs younger 
than 0.5 Gyr by \citet{m11,m12} because they are associated to the last episodes of star formation, while all the targets discussed here 
belong to the RGB, thus they are older than $\sim$1-2 Gyr.}.

The [O/Fe] ratio appears systematically lower than those measured in the Galaxy and basically consistent with those 
by \citet{pompeia08}, even if they provide a few of measures for this abundance ratio.
The [Mg/Fe] ratio turns out to be subsolar in the entire metallicity range and ever lower than the Milky Way stars; also, 
our targets show lower Mg abundances with respect to those measured by \citet{pompeia08} and 
we attribute such a discrepancy to the different values of oscillator strengths and 
damping constants for the Mg line at 5711 $\rm\mathring{A}$ employing in the two analysis. 
Finally, the [Si/Fe] ratio seems to be barely consistent with the Galactic stars, despite a larger dispersion.

Figure~\ref{alpha} shows the behaviour of the average [$\alpha$/Fe] ratio, obtained by averaging the abundances of O, Mg and Si, 
as a function of [Fe/H]. We compared our targets with the [$\alpha$/Fe] ratios of the other samples calculated 
by averaging the available $\alpha$-element abundance ratios; only for the old LMC GCs we excluded O and Mg, because  
of the intrinsic star-to-star variations of these elements due to the self-enrichment process occurred in the early stages 
of these clusters (at variance to the intermediate-age LMC GCs  where intrinsic spread in O and Mg content are not detected).
The overall trend of the [$\alpha$/Fe] ratio with the [Fe/H] abundance shows a decrease at
increasing metallicity. The most metal-poor stars with [Fe/H] $<$ --0.7 dex show an [$\alpha$/Fe] ratios larger than solar value, 
while stars with [Fe/H] $>$ --0.7 dex show [$\alpha$/Fe] ratio from solar to sub-solar values. For a given [Fe/H], the [$\alpha$/Fe] 
ratio measured in the LMC appears systematically lower than those measured in the Galaxy.

A substantial agreement of our abundances was generally found with the values measured in dSphs, in particular with 
the abundance ratios observed in the metal-poor component of the Sagittarius dwarf galaxy \citep{sbordone07}: 
such a similarity of $\alpha$-elements, metallicity distribution and 
fraction of metal-poor stars has been already suggested by \citet{monaco05}.


\section{Discussion and Conclusions}
 
We determined kinematic and chemical properties for 89 giant stars members of the disk of the LMC. 
This sample increases significantly the number of stars analysed so far through high-resolution 
spectroscopy (the largest sample observed by \citealt{pompeia08} includes 59 stars).

\begin{figure}[]
\plottwo{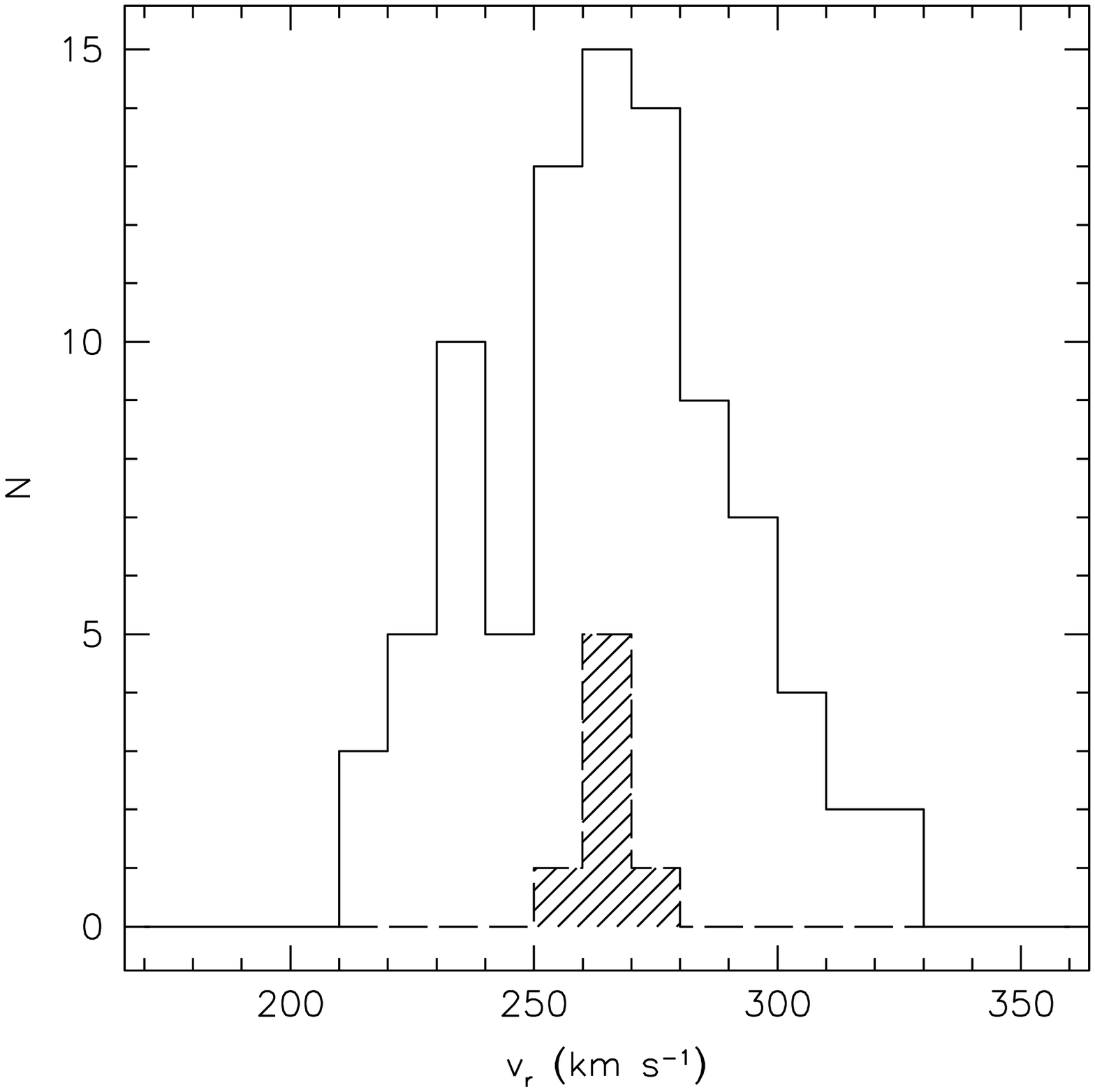}{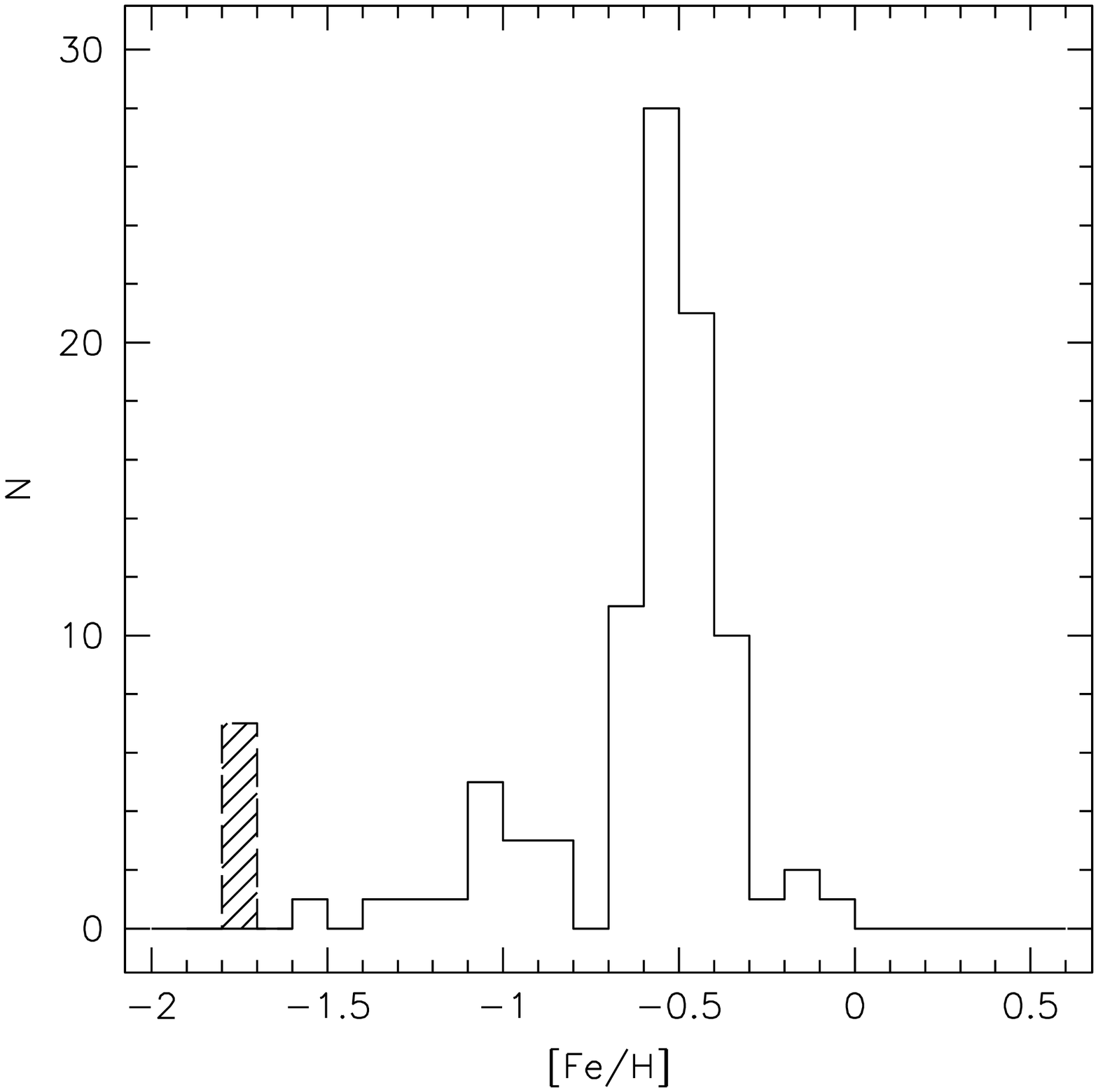}
\caption{The heliocentric radial velocity distribution (left panel) and the metallicity distribution 
(right panel) of our sample of LMC disk stars. The shaded histograms show the distribution of the 
7 member stars of NGC~1786 \citep{m09, m10}.}
\label{histogram}
\end{figure}

\begin{figure}[]
\plotone{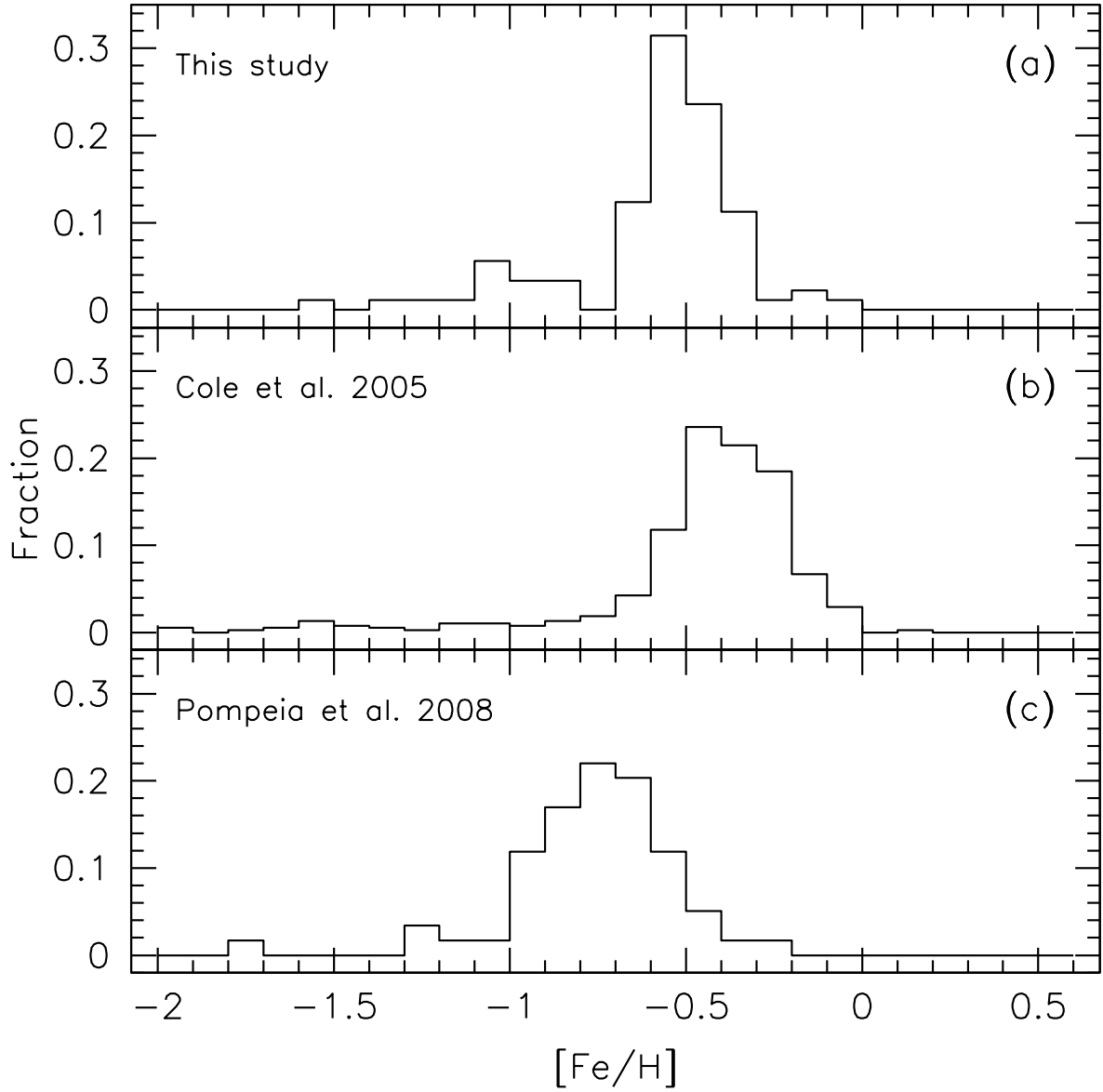}
\caption{Comparison of our normalized metallicity distribution (a) with that one by \citet{cole05} (b) 
and \citet{pompeia08} (c).}
\label{mult_hist}
\end{figure}

\begin{figure}[]
\epsscale{0.6}  
\plotone{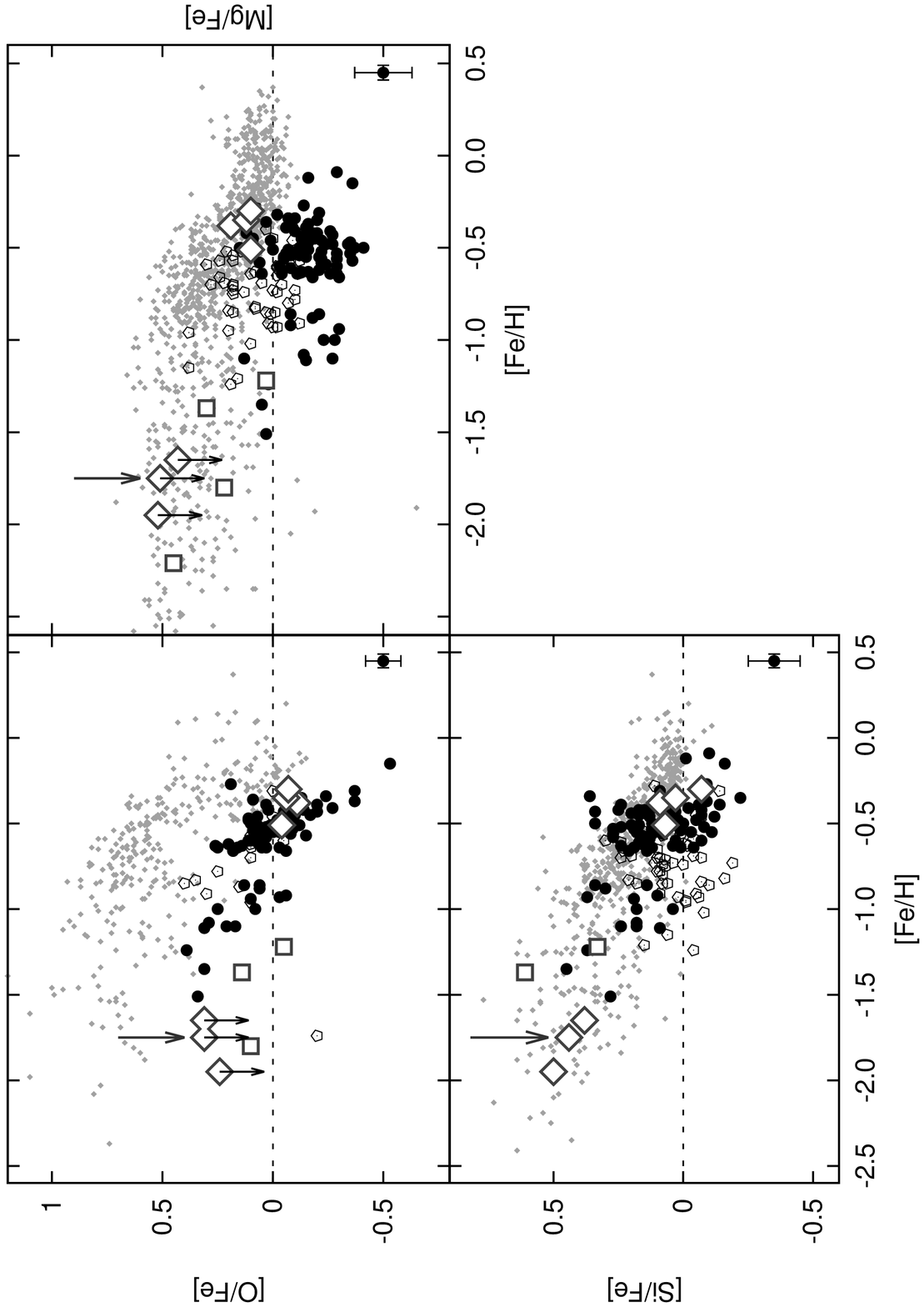}
\caption{Behaviour of the [O/Fe], [Mg/Fe] and [Si/Fe] ratio as a function of [Fe/H]. Black dots represent our targets, 
the small gray diamonds are the Galaxy data (\citealt{venn04}, \citealt{reddy06}), the black pentagons are the LMC 
disk \citep{pompeia08}, the edged white squares and diamonds are the old \citep{johnson06,m09,m10} and 
intermediate-age LMC GCs \citep{m08}, respectively.
The [Mg/Fe] abundance ratios from \cite{pompeia08} were corrected by a factor of $\simeq$ --0.11 dex to take 
into account the effect of different oscillator strength for the line at 5711 $\rm\mathring{A}$ used in both analysis. 
The average [O/Fe] and [Mg/Fe] ratios of the three old and metal poor GCs from \cite{m09} was obtained averaging the abundances 
of the stars with greater values only, in order to avoid the effects of anticorrelations. In this case the tiny black arrows 
indicate them as an ``upper limit''. The dark-gray arrow marks the position of NGC~1786. Dashed lines mark the solar value. 
The error bars in the bottom right corner indicate the cumulative (EWs + atmospheric parameters) uncertainties.}
\label{omgsi}
\end{figure}

The derived metallicity distribution is dominated by LMC-R, a metal-rich and narrow component peaked at [Fe/H] = --0.48 dex, 
with a secondary component LMC-P peaked at [Fe/H] = $\sim$ --1.0 dex and reaching [Fe/H] = $\sim$ --1.5 dex (right panel of 
Fig.~\ref{histogram}).
As shown in Figure~\ref{feh_vrad}, the two components show similar kinematic properties, with similar 
average $\langle v_{r}\rangle$ (265.9 km s$^{-1}$ for LMC-R and 261 km s$^{-1}$ for LMC-P) and with 
a small increase of the velocity dispersion decreasing the metallicity 
($\sigma_{v}$ = 25 km s$^{-1}$ and $\sigma_{v}$ = 29.7 km s$^{-1}$, respectively).

The kinematic and abundance distributions can be now compared to those observed in the stellar population of the GC NGC~1786. 
All the stars of our sample belong to the RGB, hence their age can range from $\sim$ 1-2 Gyr up to $\sim$ 12-13 Gyr, thus excluding very recent (last 500 Myr) burst of star formation. 
At variance to the stars belonging to a stellar cluster, the determination of the age for field stars through 
isochrone fitting is a dangerous and uncertain technique, because the position of a RGB star in the Color-Magnitude Diagram is 
weakly sensitive to the age and highly affected by uncertainties in the color excess, evolutive mass and distance. 
Even if precise ages for each target star cannot be determined, we can draw the timeline of the chemical evolution in this 
region of the LMC, by comparing our results with the recent simulations by \citet{bekki05} and the SFH inferred by \citet{smecker02}, 
\citet{harris09} and \citet{rubele12} in different regions of the LMC; 
unfortunately, there are no determinations of SFH in the region of our targets.

(1)~NGC~1786 is an old LMC GC generated during the first burst of SF occurred $\sim$ 12--13 Gyr ago. Following \citet{bekki05} a loose 
stellar halo of old stars with a velocity dispersion of about $\sim$ 40 km s$^{-1}$ and a broad metallicity distribution is also expected 
to be formed during that event.

It is interesting to check whether some of the stars observed in our sample could belong to NGC~1786 or to the LMC halo. Note that \citet{m09,m10} 
found for NGC~1786 $\langle v_{r} \rangle$ = 264.3 km s$^{-1}$ ($\sigma_{v}$ = 5.7 km s$^{-1}$), [Fe/H] = --1.75 dex and [$\alpha$/Fe] = 0.37 dex. 
Figure~\ref{feh_vrad} shows that none of the observed stars has kinematic and chemical properties compatible with the population of NGC~1786. 
In particular, the observed sample has a systematically higher [Fe/H] ($\gtrsim$ --1.0 dex) and a lower [$\alpha$/Fe] ($\lesssim$ 0.20 dex), thus demonstrating that even LMC-P did not originated in the cluster.

This evidence could be used to put some constraints on the presence of old metal-poor $\alpha$-enhanced stars in the sampled LMC disk field, 
which turns out to be less then $\sim$ 1$\%$. Our outcome provides a substantial confirmation of the results by \citet{cole05},
who found $\sim$ 3$\%$ of their sample with [Fe/H] $<$ --1.5 dex (but no measurements of [$\alpha$/Fe] ratio are available) and by \citet{pompeia08}, 
who found $\sim$ 2$\%$, 
although the abundance patterns are incompatible with old LMC GCs.
Since these samples cover different regions of the galaxy, the measured
similar fraction of metal-poor stars in these three samples seems to suggest that the old, metal-poor component 
is distributed in a quite homogeneous way along the LMC disk.
 
(2)~After the initial burst, the LMC has undergone a long period characterized by a continuous SF with low efficiency. 
The occurence of this quiescent period in the LMC field is clearly visible in the SFH provided 
by \citet{smecker02}, \citet{harris09} and \citet{rubele12}, where a prolonged phase with very low or lacking SF is appreciable between 
$\sim$ 4-5 Gyr and $\sim$ 12 Gyr ago. This quiescent period is predicted also by the theoretical models proposed by \citet{bekki05} 
that derived a typical SF efficiency in this age range of $\sim$ 0.1 M$_{\odot}$ yr$^{-1}$.

During this period, the LMC evolved in isolation, without gravitational interactions with the Galaxy and the SMC \citep{bekki05, besla07}. 
LMC-P likely formed during this period, as also  
suggested by the lower (by $\sim$ 0.2--0.3 dex) [$\alpha$/Fe] ratio with respect to 
the Galactic values at similar metallicity, which indicate that during this long period mainly SNeIa contribute to the gas enrichment.

If this scenario is correct, these stars were born during the period in which no GC formed (the so-called   {\sl Age Gap}, \citealt{rich01, bekki04}),
hence they are unique tracers of the LMC chemical evolution history between $\sim$ 12 and 3 Gyr ago.

We also notice that the LMC-P fraction ($\sim$ 16$\%$) is larger than that ($\lesssim$ 10$\%$) found by \citet{cole05}. 
This is probably due to different location of the two fields. The sample analysed by \cite{cole05} is located around the central bar (at variance 
with our targets, located in the inner disk at $\sim$ 1.8 kpc from the LMC center). Infact, \cite{bekki05} suggest that the central bar formed 
in the last 3--4 Gyr with a marginal fraction of metal-poor stars in its stellar content.

\begin{figure}[]
\epsscale{0.7}  
\plotone{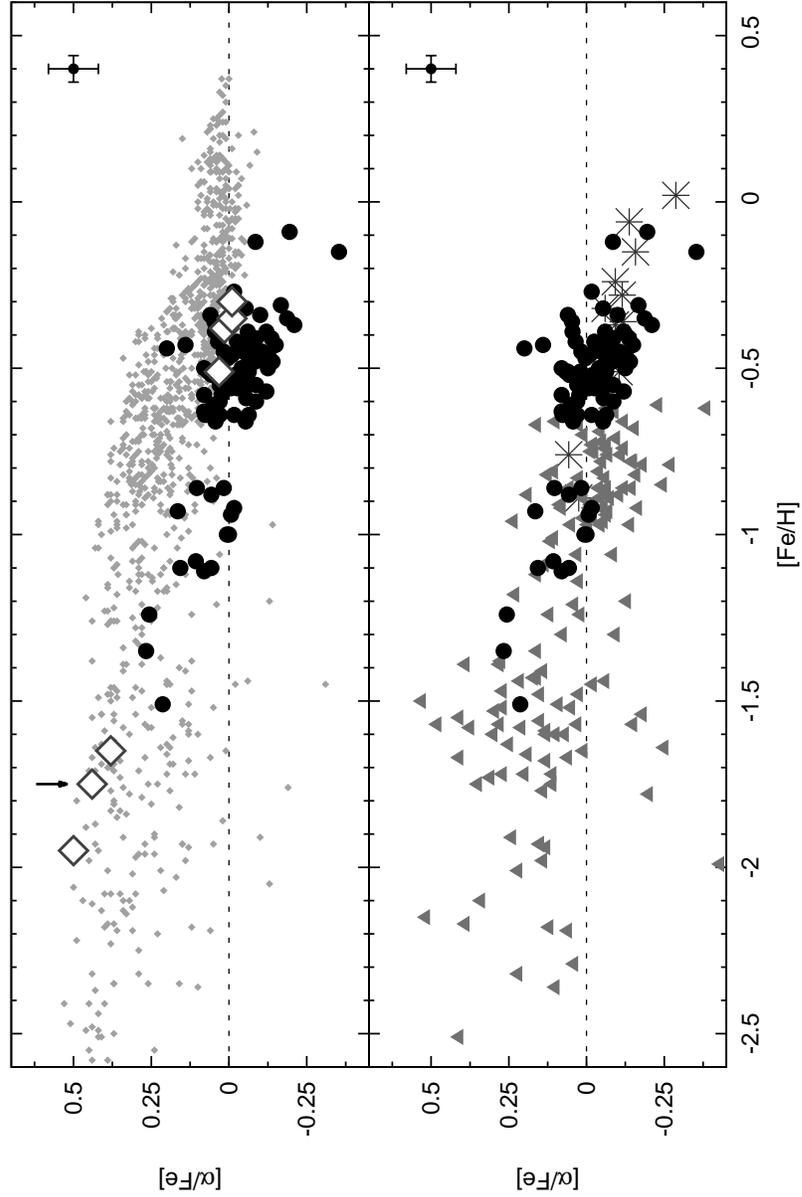}
\caption{Behaviour of the average [$\alpha$/Fe] ratio as a function of [Fe/H]. Same symbols of Figure~\ref{omgsi}. 
At variance with the intermediate-age GCs, the [$\alpha$/Fe] ratio of the three old and metal poor GCs 
includes [Si/Fe] only, in order to avoid the effect due to the intrinsic dispersion observed in the O and Mg abundances. Dashed lines 
mark the solar value. The error bars in the top right corner indicates the average uncertainty of iron and 
``$\alpha$-elements'' abundances. 
Upper panel shows the comparison with the Galactic stars, while bottom panel shows the comparison 
with stars in the Sagittarius dwarf galaxy (asterisks) and in the nearby dwarf galaxies (grey triangles; 
the plotted samples include the data by \citet{shetrone01} and \citet{shetrone03} for Draco, Sextant, Ursa Minor, Sculptor, 
Fornax, Carina and Leo~I, \citet{letarte} for 
Fornax and \citet{lemasle} and \citet{venn12} for Carina).}
\label{alpha}
\end{figure}

\begin{figure}[]
\epsscale{0.8} 
\plotone{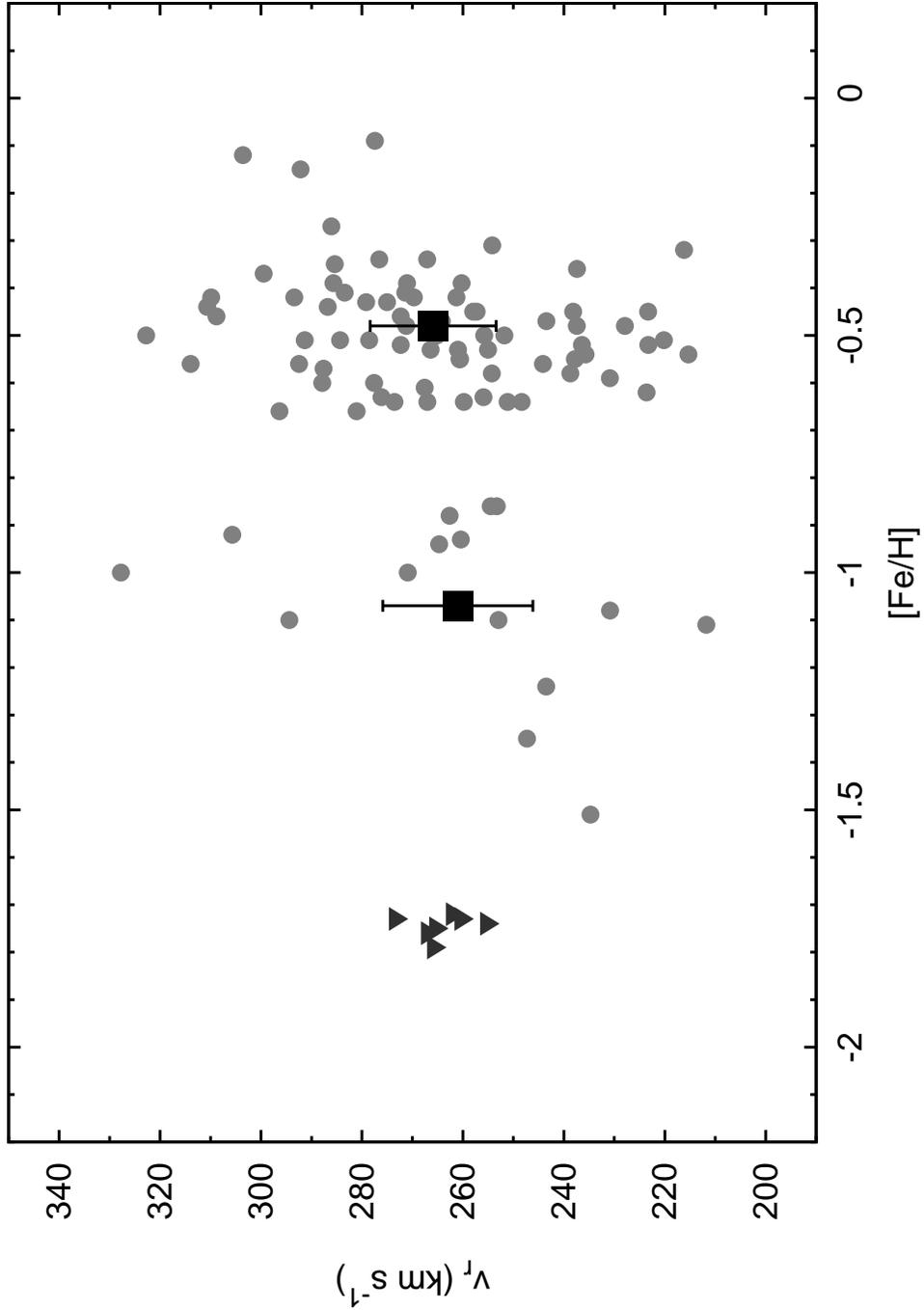}
\caption{The distribution of our targets (gray points) in the $v_{r}$ -- [Fe/H] plane with of the 7 member stars of 
NGC~1786 (black triangles) by \citet{m09, m10}. The black squares represent the average radial velocities 
and metallicities of the two components.}
\label{feh_vrad}
\end{figure}

Also, it is interesting to notice that the [$\alpha$/Fe] ratio of this stellar component well resembles the mean locus 
defined by the dSph stars (grey triangles in the lower panel of Fig.~\ref{alpha}, but see also Fig.11 in \citet{tolstoy}), suggesting similar chemical enrichment histories.

(3)~LMC-R formed during the relevant burst of SF occurred in the last few Gyr probably due to the tidal capture of the SMC by the LMC. 
Numerical simulations \citep{bekki05} predict that first close encounter between the two Clouds occurred $\sim$ 3--4 Gyr ago, when the LMC and 
SMC become a gravitational bound system. This strong interaction triggered the SFR up to $\sim$ 0.4 M$_{\odot}$ yr$^{-1}$. The occurrence of 
this SF enhancement has been confirmed by \citet{harris09} by using Color-Magnitude Diagrams of different regions of the LMC.
However, the binary system status of Magellanic Clouds and its timescales is still a matter of debate since there is no
unambiguous consensus between dynamical simulation and photometric studies.
For instance, \citet{rubele12} analysed new photometric dataset from the Vista Magellanic Survey, finding an enhancement of star formation 
with a peak at $\sim$ 2 Gyr and in some regions a second peak at $\sim$ 5 Gyr and interpreted as the epoch of tidal interaction between 
the LMC and the Galaxy.
Conversely, \citet{besla07} suggested that the LMC-SMC system is on its first close passage about the Galaxy, being entered in the
Milky Way virial radius only in the last 1-3 Gyr.

LMC-R distribution is very narrow ($\sigma_{\rm[Fe/H]}$=~0.13 dex), suggesting that the episode of SF associated to the first close encounter
between the two Clouds has been very efficient (in order to produce the majority of the LMC disk stars) and fast, because the stars have not 
had time to furtherly enrich in iron.
These stars show [$\alpha$/Fe] abundance ratios that are close to the solar value, in nice agreement with the values 
measured in the intermediate-age GCs \citep{m08} with age of 1--3 Gyr, as shown in Figure~\ref{alpha}. 

The observed [$\alpha$/Fe] ratios agree with those derived in the Sagittarius dSph giants (asterisks in lower panel of Fig.~\ref{alpha}), 
while the nearby dSphs studied so far do not reach such metallicity. The most metal-rich stars discussed by \citet{letarte} in Fornax show [Fe/H]$\sim$--0.6 dex, consistent with the metal-poor edge of the LMC-R population. These evidences 
(the similarity with Sgr and the difference with the nearby dSphs) seem to confirm that the recent SFH of the LMC has been 
quite complex and turbulent, and the stars of the LMC-R population (that dominates the metallicity distribution of our 
sample) are the products of the tidal interactions occurring among LMC, SMC and our Galaxy.

The intrinsic dispersion of each individual abundance ratio of the LMC-R 
population suggests a high degree of homogeneity, pointing out that an 
efficient and homogeneous mixing of the SN ejecta occurs. We note that the 
dispersion of the [$\alpha$/Fe] ratios in these stars is larger than the range 
of values covered by the intermediate-age GCs.

Thus, the scatter in [$\alpha$/Fe] abundances of the LMC-R stars reflects the larger 
age range in which these stars formed. However, we find a general agree 
between the values observed in the field and GC stars in this metallicity range, 
suggesting in any case a common origin of these stars from similar burst of SF.

It is important to recall that there is an offset between the age of the onset 
of the SF in the LMC field and that of the GC formation. In fact, if the 
SF in the field started about 4-5 Gyr ago (the precise age depends on the location in the disk and 
the magnitude of the tidal interactions with the SMC), the GC formation restarts 
about 2 Gyr ago, as suggested also by \citet{bekki05}.
This difference is confirmed by the measured ages in the intermediate-ages GCs \citep[see e.g.][]{rich01,m07a,m07b}
and the lack of globulars in the age range between 2 and 5 Gyr. 
Furthermore, the lower metallicity edge of LMC-R stars ([Fe/H] $\simeq$ --0.7 dex),
if compared to that of GCs, seems to support that the SF of field stars becomes
efficient slightly before.

Recently, \citet{tsujiomoto12} discussed the unusual low [Mg/Fe] (--0.9 dex) ratio measured by \citet{colucci12} in the intermediate-age, 
metal-rich GC NGC~1718, suggesting that this cluster and a fraction of LMC field stars were formed from the metal-poor gas acquired by infall 
from the SMC about 1-2 Gyr ago (the so-called {\sl Magellanic Squall}, \citealt{bekki07}).
In our sample we did not detect metal-rich stars characterized by such low Mg or $\alpha$-depletion, pointing out that the stars formed from 
metal-poor gas accreted by the SMC are (if any) a negligible ($<$1$\%$) fraction of the LMC disk.

\acknowledgements  
The authors warmly thank the anonymous referee for his/her suggestions in improving the paper.


{}


\begin{deluxetable}{rrrccc}
\tablewidth{0pc}
\tablecolumns{6}
\tiny
\tablecaption{Identification number, right ascension, declination and J, H, K magnitudes. The entire table is available in the online version.}
\tablehead{\colhead{ID} & \colhead{R.A. (J2000)} & \colhead{Dec. (J2000)} & \colhead{J} & \colhead{H} 
& \colhead{K} \\
& & & \colhead{(mag)} & \colhead{(mag)} & \colhead{(mag)} }
\startdata
 & & & & & \\
  353  &  74.8429031  &  -67.7529373  &  14.55  &  13.80  &  13.61  \\
 1415  &  74.7569885  &  -67.7277374  &  14.52  &  13.75  &  13.57  \\
 1593  &  74.7403107  &  -67.7479858  &  14.34  &  13.56  &  13.36  \\
 1954  &  74.6953583  &  -67.7596817  &  14.11  &  13.28  &  13.12  \\
 1980  &  74.6899948  &  -67.7497635  &  14.52  &  13.74  &  13.58  \\
 2353  &  74.7275314  &  -67.7423553  &  12.56  &  11.87  &  11.71  \\
 2363  &  74.7172394  &  -67.7453918  &  14.67  &  13.97  &  13.82  \\
 2376  &  74.7086487  &  -67.7528458  &  14.62  &  13.93  &  13.81  \\
90012  &  74.2391357  &  -67.7102280  &  13.91  &  13.05  &  12.93  \\
90119  &  74.5899658  &  -67.9048157  &  14.27  &  13.38  &  13.23  \\
\enddata
\label{taba}
\end{deluxetable}

\begin{deluxetable}{cccccc}
\tablewidth{0pc}
\tablecolumns{6}
\tiny
\tablecaption{Wavelength, element, oscillator strength, excitation potential and reference source of adopted line list.}
\tablehead{\colhead{$\rm\lambda$} & \colhead{El.} & \colhead{log~gf} &  accuracy &\colhead{E.P.} & \colhead{Ref.} \\
\colhead{($\rm\mathring{A}$)} & & &  & \colhead{(eV)} & }
\footnotesize
\startdata
   5586.756   &  FeI   &  -0.144  & $<$10\% &  3.37  &  \citet{fw06}  \\
   5607.664   &  FeI   &  -2.270  & $<$50\% &  4.15  &  \citet{fmw}   \\    
   5611.356   &  FeI   &  -2.990  & $<$50\% &  3.64  &  \citet{fmw}   \\    
   5618.632   &  FeI   &  -1.276  & $<$18\% &  4.21  &  \citet{fw06}  \\
   5619.225   &  FeI   &  -3.270  & $<$50\% &  3.69  &  \citet{fmw}   \\    
   5619.595   &  FeI   &  -1.670  & $>$50\% &  4.39  &  \citet{fw06}  \\
   5624.022   &  FeI   &  -1.140  & $<$15\% &  4.39  &  SUN  \\
   5633.946   &  FeI   &  -0.320  & $>$50\% &  4.99  &  \citet{fw06}  \\
   5636.696   &  FeI   &  -2.560  & $>$50\% &  3.64  &  \citet{fw06}  \\
   5638.262   &  FeI   &  -0.840  & $<$50\% &  4.22  &  \citet{fw06}  \\
   5640.307   &  FeI   &  -1.700  & $<$15\% &  4.64  &  SUN  \\
   5646.684   &  FeI   &  -2.500  & $<$50\% &  4.26  &  \citet{fmw}   \\    
   5650.690   &  FeI   &  -0.960  & $>$50\% &  5.09  &  \citet{fw06}  \\
   5651.469   &  FeI   &  -2.000  & $<$50\% &  4.47  &  \citet{fmw}   \\    
   5652.318   &  FeI   &  -1.920  & $>$50\% &  4.26  &  \citet{fw06}  \\
   5653.867   &  FeI   &  -1.610  & $>$50\% &  4.39  &  \citet{fw06}  \\
   5661.021   &  FeI   &  -2.430  & $<$50\% &  4.58  &  \citet{fmw}   \\   
   5661.345   &  FeI   &  -1.756  & $<$10\%  &  4.28  &  \citet{fw06}  \\
   5662.516   &  FeI   &  -0.573  & $<$10\%  &  4.18  &  \citet{fw06}  \\
   5677.684   &  FeI   &  -2.700  & $<$50\% &  4.10  &  \citet{fmw}   \\   
   5678.601   &  FeI   &  -4.670  & $<$50\% &  2.42  &  \citet{fmw}   \\   
   5680.240   &  FeI   &  -2.540  & $>$50\% &  4.19  &  \citet{fw06}  \\
   5691.497   &  FeI   &  -1.490  & $>$50\% &  4.30  &  \citet{fw06}  \\
   5693.640   &  FeI   &  -0.680  & $<$15\% &  4.96  &  SUN  \\
   5704.733   &  FeI   &  -1.250  & $<$15\% &  5.03  &  SUN  \\
   5705.465   &  FeI   &  -1.355  & $<$10\%  &  4.30  &  \citet{fw06}  \\
   5714.551   &  FeI   &  -1.770  & $<$15\% &  5.09  &  SUN  \\  
   5720.886   &  FeI   &  -1.950  & $<$50\% &  4.55  &  \citet{fmw}   \\    
   5731.762   &  FeI   &  -1.270  & $<$50\% &  4.26  &  \citet{fw06}  \\
   5732.296   &  FeI   &  -1.560  & $<$25\% &  4.99  &  \citet{fmw}   \\   
   5741.848   &  FeI   &  -1.670  & $<$25\%  &  4.26  &  \citet{fw06}  \\
   5760.345   &  FeI   &  -2.440  & $>$50\% &  3.64  &  \citet{fw06}  \\
   5767.972   &  FeI   &  -3.200  & $<$15\%  &  4.29  &  SUN  \\
   5775.081   &  FeI   &  -1.298  & $<$18\%  &  4.22  &  \citet{fw06}  \\
   5776.224   &  FeI   &  -3.400  & $<$15\% &  3.69  &  SUN  \\
   5778.453   &  FeI   &  -3.430  & $<$25\% &  2.59  &  \citet{fw06}  \\
   5793.915   &  FeI   &  -1.660  & $>$50\% &  4.22  &  \citet{fw06}  \\
   5805.757   &  FeI   &  -1.590  & $<$25\% &  5.03  &  \citet{fmw}   \\    
   5806.725   &  FeI   &  -1.030  & $<$50\% &  4.61  &  \citet{fw06}  \\
   5811.915   &  FeI   &  -2.430  & $<$50\% &  4.14  &  \citet{fmw}   \\   
   6120.249   &  FeI   &  -5.970  & $<$7\% &  0.91  &  \citet{fw06}  \\
   6151.618   &  FeI   &  -3.299  & $<$10\%  &  2.18  &  \citet{fw06}  \\
   6165.360   &  FeI   &  -1.474  & $<$18\% &  4.14  &  \citet{fw06}  \\
   6187.989   &  FeI   &  -1.670  & $>$50\% &  3.94  &  \citet{fw06}  \\
   6200.313   &  FeI   &  -2.437  & $<$18\% &  2.61  &  \citet{fw06}  \\
   6226.734   &  FeI   &  -2.220  & $<$50\% &  3.88  &  \citet{fmw}   \\      
   6246.319   &  FeI   &  -0.877  & $<$10\%  &  3.60  &  \citet{fw06}  \\
   6252.555   &  FeI   &  -1.687  & $<$10\%  &  2.40  &  \citet{fw06}  \\
   6322.686   &  FeI   &  -2.426  & $<$18\% &  2.59  &  \citet{fw06}  \\
   6330.849   &  FeI   &  -1.190  & $<$15\%  &  4.73  &  SUN  \\
   6335.330   &  FeI   &  -2.177  & $<$18\% &  2.20  &  \citet{fw06}  \\
   6336.824   &  FeI   &  -0.856  & $<$10\%  &  3.69  &  \citet{fw06}  \\
   6380.743   &  FeI   &  -1.376  & $<$25\% &  4.19  &  \citet{fw06}  \\
   6300.304   &  OI    &  -9.717  & $<$2\% &  0.00  &  \citet{storey}   \\
   5711.095   &  MgI   &  -1.724  & $<$10\%  &  4.35  &  NIST  \\
   5665.555   &  SiI   &  -2.040  & 20\% &  4.92  &  \citet{garz}  \\
   5666.677   &  SiI   &  -1.710  & $<$15\% &  5.62  &  SUN  \\
   5690.425   &  SiI   &  -1.870  & 20\% &  4.93  &  \citet{garz}  \\
   5701.104   &  SiI   &  -2.050  & 20\% &  4.93  &  \citet{garz}  \\
   5793.073   &  SiI   &  -2.060  & 20\% &  4.93  &  \citet{garz}  \\
   6155.144   &  SiI   &  -0.870  & $<$15\% &  5.62  &  SUN  \\
   6237.319   &  SiI   &  -1.100  & $<$15\%  &  5.61  &  SUN  \\
\enddata
\label{tabb}
\end{deluxetable}

\normalsize

\begin{deluxetable}{rcccccrc}
\tablewidth{0pc}
\tablecolumns{8}
\tiny
\tablecaption{Identification number, photometric temperature and gravity, radial velocity and spectroscopic atmospheric parameters for the stars in our sample. The entire table is available in the online version.}
\tablehead{\colhead{ID} & \colhead{$T_{\rm eff}^{\rm phot}$} & \colhead{log~$g^{\rm phot}$} & \colhead{$v_{r}$} 
& \colhead{$T_{\rm eff}^{\rm spec}$} & \colhead{log~$g^{\rm spec}$} & \colhead{[A/H]} & \colhead{$v_t^{\rm spec}$} \\
& \colhead{(K)} & \colhead{(dex)} & \colhead{(km s$^{-1}$)} & \colhead{(K)} & \colhead{(dex)} & \colhead{(dex)} & \colhead{(km s$^{-1}$)}}
\startdata
 & & & & & & & \\
  353  &  4074  &  1.12  &  269.7  &  4150  &  1.10  &  --0.50  &  1.50  \\
 1415  &  4043  &  1.10  &  223.6  &  4200  &  1.10  &  --0.50  &  1.80  \\
 1593  &  3989  &  1.01  &  254.2  &  4150  &  1.10  &  --0.50  &  1.60  \\
 1954  &  3922  &  0.91  &  223.3  &  3850  &  0.90  &    0.00  &  1.50  \\
 1980  &  4018  &  1.10  &  272.3  &  4100  &  1.10  &  --0.50  &  1.60  \\
 2353  &  4255  &  0.39  &  299.4  &  4350  &  0.40  &  --0.50  &  1.80  \\
 2363  &  4227  &  1.23  &  265.1  &  4350  &  1.30  &  --0.50  &  1.70  \\
 2376  &  4308  &  1.24  &  230.8  &  4600  &  1.30  &  --1.00  &  2.10  \\
90012  &  3906  &  0.83  &  243.5  &  4000  &  0.80  &  --1.00  &  2.00  \\
90119  &  3795  &  0.93  &  251.1  &  4450  &  1.20  &  --0.50  &  2.10  \\
\enddata
\label{tab1}
\end{deluxetable}

\begin{deluxetable}{rrrrr}
\tablewidth{0pc}
\tablecolumns{5}
\tiny
\tablecaption{Derived abundances for the stars of our sample. The entire table is available in the online version.}
\tablehead{\colhead{ID} & \colhead{[Fe/H]} & \colhead{[O/Fe]} &  \colhead{[Mg/Fe]} & \colhead{[Si/Fe]} \\
& \colhead{(dex)} & \colhead{(dex)} & \colhead{(dex)} & \colhead{(dex)} }
\startdata
 & & & & \\
  353  &  --0.42 $\pm$ 0.03  &   0.01 $\pm$ 0.05  &  --0.16 $\pm$ 0.08  &  --0.07 $\pm$ 0.06  \\
 1415  &  --0.62 $\pm$ 0.03  &   0.19 $\pm$ 0.04  &  --0.21 $\pm$ 0.02  &   0.16 $\pm$ 0.03  \\
 1593  &  --0.58 $\pm$ 0.02  &   0.04 $\pm$ 0.06  &   0.06 $\pm$ 0.09  &   0.14 $\pm$ 0.05  \\
 1954  &  --0.45 $\pm$ 0.03  &  --0.18 $\pm$ 0.09  &  --0.13 $\pm$ 0.10  &   0.17 $\pm$ 0.05  \\
 1980  &  --0.46 $\pm$ 0.02  &   0.06 $\pm$ 0.04  &  --0.22 $\pm$ 0.10  &   0.13 $\pm$ 0.02  \\
 2353  &  --0.37 $\pm$ 0.02  &  --0.38 $\pm$ 0.08  &  --0.16 $\pm$ 0.09  &  --0.09 $\pm$ 0.03  \\
 2363  &  --0.50 $\pm$ 0.03  &  --0.03 $\pm$ 0.07  &  --0.12 $\pm$ 0.10  &   0.03 $\pm$ 0.04  \\
 2376  &  --1.08 $\pm$ 0.03  &   0.28 $\pm$ 0.05  &  --0.14 $\pm$ 0.11  &   0.18 $\pm$ 0.06  \\
90012  &  --1.24 $\pm$ 0.02  &   0.38 $\pm$ 0.09  &   0.02 $\pm$ 0.07  &   0.37 $\pm$ 0.07  \\
90119  &  --0.64 $\pm$ 0.02  &   0.24 $\pm$ 0.09  &  --0.11 $\pm$ 0.13  &  --0.04 $\pm$ 0.05  \\
\enddata
\tablecomments{The adopted solar values are 7.50, 8.76, 7.58 and 7.55 for Fe, O, Mg and Si.}
\label{tab2}
\end{deluxetable}

\begin{deluxetable}{lccccc}
\tablewidth{0pc}
\tablecolumns{6}
\tiny
\tablecaption{Abundance uncertainties for the star 91969. The second column is the total uncertainty 
estimated by following the prescriptions by \citet{cayrel04}. The other columns show the variations in abundance 
due to the variation of one only parameter, while the last column is the sum in quadrature of these terms, 
without taking into account the covariance terms.}
\tablehead{\colhead{Element} & \colhead{Parameters} & \colhead{$\delta T_{eff}$} 
 & \colhead{$\delta$log~g} & \colhead{$\delta v_{t}$} & \colhead{Quadrature}\\
\colhead{} & \colhead{uncertainty} & \colhead{$\pm$100K} & \colhead{$\pm$0.1} & \colhead{$\pm$0.1 km s$^{-1}$} \\
\colhead{} & \colhead{(dex)} & \colhead{(dex)} & \colhead{(dex)} & \colhead{(dex)}& \colhead{(dex)} }
\startdata
Fe &  $\pm$0.01  & $\pm$0.03 & $\pm$0.02  & $\pm$0.04   &  $\pm$0.05\\
O  &  $\pm$0.02  & $\pm$0.01 & $\pm$0.05  & $\pm$0.02   &  $\pm$0.05\\
Mg &  $\pm$0.04  & $\pm$0.03 & $\pm$0.01  & $\pm$0.02   &  $\pm$0.04\\    
Si &  $\pm$0.06  & $\pm$0.07 & $\pm$0.03  & $\pm$0.02   &  $\pm$0.08\\
\enddata
\label{tab3}
\end{deluxetable}



\end{document}